\documentstyle[epsfig]{article}
\begin{document}
\title{\bf Numerical modelling of Bose-Einstein correlations}
\author{O.V.Utyuzh$^{1}$\thanks{e-mail: utyuzh@fuw.edu.pl},~
G.Wilk$^{1}$\thanks{e-mail: wilk@fuw.edu.pl} and Z.W\l odarczyk$^{2}$
\thanks{e-mail: wlod@pu.kielce.pl}\\[2ex]
$^1${\it The Andrzej So\l tan Institute for Nuclear Studies}\\
    {\it Ho\.za 69; 00-689 Warsaw, Poland}\\
$^2${\it Institute of Physics, \'Swi\c{e}tokrzyska Academy}\\
    {\it  \'Swi\c{e}tokrzyska 15; 25-405 Kielce, Poland}}
\date{\today}
\maketitle

\begin{abstract}
We propose extension of the algorithm for numerical modelling of
Bose-Einstein correlations (BEC), which was presented some time ago
in the literature. It is formulated on quantum statistical level for
a single event and uses the fact that identical particles subjected
to Bose statistics do bunch themselves, in a maximal possible way, in
the same cells in phase-space. The bunching effect is in our case
obtained in novel  way allowing for broad applications and fast
numerical calculations. First comparison with $e^+e^-$ annihilations
data performed by using simple cascade hadronization model is very
encouraging.\\  

PACS numbers: 25.75.Gz 12.40.Ee 03.65.-w 05.30.Jp\\
{\it Keywords:} Bose-Einstein correlations; Statistical models;
Fluctuations \\

\end{abstract}


Bose-Einstein correlations (BEC) between identical bosons are since
long time of special interest because of their ability to provide
space-time information about multiparticle production processes
\cite{BEC}. This is particulary true in searches for a proper
dynamical evolution of heavy ion collisions (QGP) \cite{QGP}.
However, such processes, because of their complexity, must be
modelled by means of Monte Carlo event generators \cite{GEN},
probabilistic structure of which prevents {\it a priori} the genuine
BEC (which are of purely quantum statistical origin). One can only
attempt to {\it model} BEC in some way aiming to reproduce the
two-particle correlation function measured experimentally and
defined, for example, as ratio of the two-particle distributions to
the product of single-particle distributions: 
\begin{equation}
C_2(Q=\vert p_i - p_j\vert )\, =\,
             \frac{N_2(p_i,p_j)}{N_1(p_i)\, N_1(p_j)} . \label{eq:C2}
\end{equation}
This is done either by changing the original output of these
generators by artificially bunching in phase-space (in a suitable
way) the finally produced identical particles \cite{LS,FW} or by
constructing generator, which allows to account from the very
beginning for the bosonic character of produced secondaries 
\cite{OMT}\footnote{The specific approaches proposed for LUND model 
\cite{AR} and the afterburner method discussed in \cite{AFTER},
which we shall not discussed here, belong to first cathegory.}. In
former case the simplest approach is to shift (in each event) momenta
of adjacent like-charged particles in such a way as to get
desired $C_2(Q)$ \cite{LS} and to correct afterwards for the
energy-momentum imbalance introduced this way. Much more physical is
the method developed in \cite{FW}, which screens all events against
the possible amount of bunching they are already showning and counts
them as many times as necessary to get desirable
$C_2(Q)$\footnote{Technically this is realised by multiplying each 
event by a special weight calculated using the output provided by
event generator.}. The original energy-momentum balance remains in
this case intact whereas the original single particle distributions
are  changed (this fact can be corrected by running again generator
with suitably modified input parameters). The size parameters
occuring in weights bear no direct resemblance to the size parameter
$R$ obtained by directly fitting data on $C_2(Q)$ in eq.(\ref{eq:C2})
by, for example, simple gaussian form: $C_2(Q) = c\left[1 + \lambda
\exp\left( - R^2 Q^2\right) \right]$ (where $c$ is normalization
constant, $\lambda$ the so called chaoticity and $R$ the size
parameter). They represent instead the corresponding correlation
lengths between the like particles \cite{BEC}.\\ 

The approach proposed in \cite{OMT} represents different philosophy
of getting desired bunching. Here one uses specific generator, which 
groups (already on a single event level) bosonic particles of the
same charge in a given cell in phase space according to Bose-Einstein
distribution\footnote{Similar concept of elementary emmiting cells
has been also proposed in \cite{BSWW}.},   
\begin{equation}
P(E_i) \sim  \exp \left[ \frac{n_i \left(\mu - E_i\right)}{T}\right]. 
\label{eq:OMT}
\end{equation}
Here $n_i$ is their multiplicity and $E_i$ energy, the
energy-momentum and charge conservations are strictly imposed by
means of the information theory concept of maximazing suitable
information entropy. The parameters $T$ and $\mu$ are
therefore two lagrange multipliers with values fixed by the
energy-momentum and charge conservation laws, respectively. Such
distribution is typical example of nonstatistical fluctuations
present in the hadronizing source. With only one additional parameter
$\delta y$, which denotes the size of the elementary emitting cell
in phase-space (in \cite{OMT} it means in rapidity), one gets at the
same time both the correct BEC pattern (i.e., correlations) and
fluctuations (as characterized by intermittency) \cite{OMT}. This is
very strong advantage of this model, which is so far the only example
of hadronization model, in which Bose-Einstein statistics is not only
included from the very beginning on a single event level, but it is
also properly used in getting final secondaries. In all other
approaches at least one of the above elements is missing. The
shortcoming of method \cite{OMT} are numerical difficulties to keep
the energy-momentum conservation as exact as possible and limitation
to the specific event generator only.\\

In the present work we propose generalization of this approach making
it applicable to other generators. Namely, following the same
reasoning as in \cite{OMT,BSWW}, we propose different method of
introducing desired bunching. In \cite{OMT} the generator itself
provided particles satisfying Bose statistics (in the sense mentioned
above). In the general case one has to find the possible bosonic
configurations of secondaries existing already among the produced
particles. The point is that nonstatistical fluctuations present in
each event generator result in a nonuniform (bunched) distributions
of particles produced in a given event in momentum space. They
resemble Bose distribution provided by the statistical event
generator of \cite{OMT}, the only difference being that particles in
such bunches usually have different charges allocated to them by
event generator, whereas in \cite{OMT} particles were of the same
charge. We propose therefore to maximally equalize charges of
particles in such bunches (to the extent limited only by the overall
charge conservation)  superseeding the initial charge allocation
provided by event generator (keeping only intact the total number of
particles of each charge it gives). This is supposed to be done in
each single event. Both the  original energy-momentum pattern
provided by event generator and all inclusive single particle
distributions are left intact\footnote{The necessary changes to event
generator used introduced by such procedure will be discussed later.
What we propose here is to resign from the part of the information
provided by event generator concerning the charge allocation to
produced particles. This can be regarded as introduction of quantum
mechanical element of uncertainty to the otherwise classical 
scheme of generator used (however, it differs completely from the
usual attempts to introduce quantum mechanical effects discussed in 
\cite{QUANT}).}.\\ 

It is instructive to look at this problem from yet another point of
view. Namely, it can be perceived as an attempt (cf. \cite{F}) to
model correlations of fluctuations present in the system, as given
by:     
\begin{eqnarray}
\langle n_1 n_2\rangle &=&  \langle n_1\rangle \langle n_2\rangle
                     + \langle \left(n_1 - \langle n_1\rangle\right)
               \left(n_2 - \langle n_2\rangle\right)\rangle \nonumber\\
                     &=& \langle n_1\rangle \langle n_2\rangle
                         + \rho \sigma(n_1)\sigma(n_2). \label{eq:COV}
\end{eqnarray}
Here $\sigma(n)$ is dispersion of the multiplicity distribution $P(n)$
and $\rho$ is the correlation coefficient depending on the type of
particles produced: $\rho = +1,-1,0$ for bosons, fermions and
Boltzmann statistics, respectively. The proposed algorithm should 
provide us with $C_2(Q)$, which is in fact a measure of correlation
of fluctuations because
\begin{equation}
C_2(Q=|p_i-p_j|) = \frac{\langle
n_i\left(p_i\right)n_j\left(p_j\right)\rangle}
  {\langle n_i\left(p_i\right)\rangle\langle
n_j\left(p_j\right)\rangle}
       = 1 + \rho \frac{\sigma\left(n_i\right)}
                             {\langle n_i\left(p_i\right)\rangle}
                        \frac{\sigma\left(n_j\right)}
                        {\langle n_j\left(p_j\right)\rangle} .
\label{eq:algor}
\end{equation}
To get $\rho > 0$ (in which we are interested here) it is enough to
select one of the produced particles, allocate to it some charge, and
then allocate (in some prescribe way) the same charge to as many
particles located near it in the phase space as possible (limited
only by the charge conservation constraint\footnote{The
energy-momentum constraint is taken care by the generator itself and
is not affected by our algorithm.}). In this way one forms a cell in
phase-space, which is occupied by particles of the same charge only.
This process should then be repeated until all particles are used and
it should be done in such way as to get geometrical (Bose-Einstein)
distribution of particles in a given cell. We stress again that this
procedure does not alter neither the original energy-momentum
distributions nor the spatio-temporal pattern of particles provided
by event generator. It only changes the charge flow pattern it
provides retaining, however, both the initial charge of the system
and its total multiplicity distribution.  Therefore this method works
only when we can resign from controlling the charge flow during
hadronization process.\\ 

The procedure of formation of such cells is controlled by a parameter
$P$ being the probability that given neighbor of the initially
selected particle  should be counted as another member of the newly
created emmiting cell in phase space. Notice that any selection
procedure leading to a geometrical particle distribution in cells (in
which case $\sigma = <n>$) results in maximization of the second term
in the eq. (\ref{eq:algor}). In particular it can be realized by the
following algorithm of allocation of charges (the $n_l = n_l^{(+)} +
n_l^{(-)} + n_l^{(0)}$ is the number of particles of different
charges provided by our event generator in the $l^{th}$ event,
$\{p_j\}$ and $\{x_i\}$ are, respectively, their energy-momenta and
spatio-temporal positions, which we keep intact):
\begin{itemize}
\item [$(1)$] The {\it SIGN} is chosen randomly from: "+", "-"or "0"
pool, with probabilities given by $p^{(+)}_l=n_l^{(+)}/n_l$,
$p^{(-)}_l = n_l^{(-)}/n_l$ and $p^{(0)}_l=n_l^{(0)}/n_l$. It is
attached to particle $(i)$ chosen randomly from the particles
produced in this event and not yet reassigned new charges.
\item[$(2)$] Distances in momenta, $\delta_{ij}(p) = \left|p_i -
p_j\right| $, between the chosen particle $(i)$ and all other
particles $(j)$ still without signs are calculated and arranged in
ascending order with $j=1$ denoting the nearest neighbor of particle
$(i)$. To each $\delta_{ij}(p)$ one assignes some probability
$P(i,j)\in (0,1)$.
\item[$(3)$] A random number $r \in (0,1)$ is selected from a uniform
distribution. If $n^{SIGN}_l > 0$, i.e., if there are still particles
of given {\it SIGN} with not reassigned charges, one checks the
particles $(j)$ in ascending order of $j$ and if $r < P(i,j)$ then
charge {\it SIGN} is assigned also to the particle $(j)$, the
original multiplicity of particles with this {\it SIGN} is reduced
by one, $n^{SIGN}_l = n^{SIGN}_l -1$, and the next particle is
selected: $(j)\Rightarrow (j+1)$. However, if  $r> P(i,j)$ or
$n^{SIGN}_l =0$ then one  returns to point $(1)$ with the updated
values of $p_l^{(+)},~p_l^{(-)}$ and $p_l^{(0)}$. Procedure finishes
when $n_l^{(+)} = n_l^{(-)} = n_l^{(0)} = 0$, in which case one
proceeds to the next event.
\end{itemize}
As can be easily checked, this algorithm results in geometrical
(Bose-Einstein) distribution of particles in the phase-space cells
formed by our procedure (with mean multiplicity $P/(1-P)$ for constant
$P(ij)=P$ case) accounting therefore for their bosonic character
(i.e., for Bose-Einstein statistics they should obey)\footnote{It is
important to realize that, because we do not restrict {\it a priori}
the number of particles which can be put in a given cell, we are
automatically getting BEC of {\it all orders} (even if we use only
two particle  checking procedure at a given step in our algorithm).
It means that $C_2(Q=0)$ calculated in such environment of the
possible multiparticle BEC can exceed $2$ (cf. \cite{FW}).}.\\

We shall illustrate now action of our algorithm on simple cascade
model of hadronization (CAS) (in its one-dimensional versions and
assuming, for simplicity, that only direct pions are
produced)\cite{CAS}. In CAS the initial mass $M$ hadronizes by 
series of well defined (albeit random) branchings ($M\rightarrow
M_1+M_2$) and is endowed with a simple spatio-temporal pattern. It
shows no traces of Bose-Einstein statistics whatsoever. However, as
can be seen in Fig. 1, when endowed with charge selection provided by
our algorithm, a clear BEC pattern emerges in $C_2(Q)$. Two kind of
choices of probabilities are shown in Fig. 1. First is constant
$P=0.75$ and $P=0.5$. It leads to a pure geometrical distribution of
number of particles allocated to a given cell and corresponds to
situation already encountered in \cite{OMT}. Its actual value is so
far a free parameter replacing, in a sense, the parameter $\delta y$
in \cite{OMT}. However, whereas in \cite{OMT} the size of emitting
cells was fixed, in our case it is fluctuating. The other is what we
call the "minimal" weight constructed from the output information
provided by CAS event generator: 
\begin{equation}
P(ij) = \exp\left[- \frac{1}{2} \delta^2_{ij}(x)\cdot\delta^2_{ij}(p)
\right]  \label{eq:CAS}
\end{equation}
where $\delta_{ij}(x) = |x_i-x_j|$ and $\delta_{ij}(p) = |p_i
- p_j|$. In this way one connects $P$ with details of hadronization
process by introducing to it a kind of overlap between particles as a
measure of probability of their bunching in a given emitting cell.\\

As we have checked out the BEC effect occuring here depends only on
the (mean) number of particles of the same charge in phase-space cell
and on the (mean) numbers of such cells. This depends on $P$, the
bigger $P$ the more particles and bigger $C_2(Q=0)$; smaller $P$
leads to the increasing number of cells, which, in turn, results in
decreasing $C_2(Q=0)$, as already noticed in \cite{BSWW}. For small
energies the number of cells decreases in natural way while their
occupation remains the same (because $P$ is the same), therefore the
corresponding $C_2(0)$ is bigger, as seen in Fig. 1. The fact that 
there is tendency to have $C_2(0) > 2$ for larger $P$ means that one
has in this case more cells with more than $2$ particles allocated to
them, i.e., it is caused by the influence of higher order BEC.
Therefore the "sizes" $R$ obtained from the exponential fits to
results in Fig. 1 (like $C_2(Q) \sim 1 + \lambda\cdot \exp(-Q\cdot
R)$ where $\lambda$ being usually called chaoticity parameter
\cite{BEC}) correspond to the sizes of the respective elementary
cells rather than to sizes of the whole hadronizing sources itself. 
For $P=0.5$ the "size" $R$ varies weakly between $0.66$ to $0.87$ fm
from $M=10$ to $100$ GeV whereas for the "minimal" weight
(\ref{eq:CAS})  it varies from $0.64$ to $0.44$ fm \footnote{This
should be contrasted with the "real" (mean) sizes of CAS sources
changing from $0.43$ fm for $M=10$ GeV to $1.82$ fm for $M=100$ GeV
\cite{CAS}.}.\\   

So far we were considering only single sources. Suppose now that
source of mass $M$ consists of a number ($n_l=2^k$) of subsources
hadronizing independently. It turns out that the resulting $C_2$'s
are very sensitive to whether in this case one applies our algorithm of
assigning charges to all particles from subsources taken together
("Split" type of sources) or to each of the subsource independently
("Indep" type of sources), cf. Fig. 2. Whereas the later case
results in the similar "sizes" $R$ (defined as before) with $C_2(Q=0)
- 1 = \lambda$ falling dramatically with increasing $k$ (roughly like
$1/2^k$, i.e., inversely with the number of subsources, $n_l$, as
expected fom \cite{BSWW}), the former leads to roughly the same
$C_2(Q=0)$ but the "size" $R$ is now increasing substantially being
equal to, respectively, $0.87$ fm, $1.29$ fm, $1.99$ fm and $3.35$ fm
for $P=0.5$ and $0.57$ fm, $3.26$ fm, $4.01$ and $5.59$ fm for the
"minimal" weight (\ref{eq:CAS}). Fig. 3 contains example of our "best
fit" to the $e^+e^-$ annihilation data on BEC by DELPHI Collaboration
\cite{DBEC} for $M=91.3$ GeV (which can be obtained only for two or
three subsources, as shown there).\\ 

To summarize: we propose a new and simple method of numerical 
modelling of BEC. It is based on reassigning charges of produced
particles in such a way as to make them look like particles
satisfying Bose statistics, conserves the energy-momenta and does not
alter the spatio-temporal pattern of events or any single particle
inclusive distribution (but it can change the distributions of,
separately, charged and neutral particles leaving, however, the total
distribution intact). It is intended to be a kind of suitable
extension of the algorithm presented in \cite{OMT}, such that can be
applied to essential any event generator in which such reassignment
of charges is possible. It amounts to the changes in physical picture
of the original generator. The example of CAS is very illustrative in
this respect. In it, at each branching vertex one has, in addition to
the energy-momentum conservation, imposed strict charge conservation
and one assumes that only $(0)\rightarrow (+-)$, $(+)\rightarrow
(+0)$ and $(-)\rightarrow (-0)$ transitions are possible. It means
that there are no multicharged vertices (i.e., vertices with multiple
charges of the same sign) in the model. However, after applying to
the finally produced particles our charge reassignment algorithm one
finds, when working the branching tree "backwards", that precisely
such vertices occur now (with charges "(++)", or "(-~-)", for
example). The total charge is, however, still conserved as are the
charges in decaying vertices (i.e., no spurious charge is being
produced because of our algorithm). It is plausible therefore that to
get BEC in an event generators it is enough to allow for cumulation
of charges of the same sign at some points of hadronization procedure
modelled by this generator. This would lead, however, to extremely
difficult numerical problem with ending such algorithms without
producing spurious multicharged particles not observed in
nature\footnote{It should be noted that possibility of using
multi(like)charged resonanses or clusters as possible source of BEC
has been recently mentioned in \cite{BUSH}. There remains problem of
their modellig, which although clearly visible in CAS model, is not
so straightforward in other approaches. However, at least in the LUND
model (or other string models) one can imagine that it could proceed
through the formation of charged (instead of neutral) dipoles, i.e.,
by allowing formation of multi(like)charged systems of opposite signs
out of vacuum when breaking the string. Because only a tiny fraction
of such processes seems to be enough in CAS, it would probably be
quite acceptable modification in the string model approach
\cite{LUND}.}.\\  

We find that value of $C_2(Q=0)$ (defining chaoticity parameter
$\lambda$) depends inversely on the number of elementary cells, in
the way already discussed in \cite{BSWW}, and that "radius" $R$
extracted from the exponential fits is practically independent on the
size of the source, provided it is a single one. In the case when it
is composed of a number of elementary sources, $R$ increases with
their number, unless they are treated independently by our algorithm.
It is because one has in this case a higher density of particles.
This results in smaller average $Q$, and this in turn leads to bigger
$R$\footnote{This feature of our model allows to understand the
increase of the extracted "size" parameter $R$ with $A$ in nuclear
collisions. That is because with increasing $A$ the number of
collided nucleons, which somehow must correspond to the number of
sources in our case, also increases. If they turn out to be of the
"Split" type, the increase of $R$ follows then naturally.}. On the
contrary, for the independently treated sources the density of
particles subjected to our algorithm does not change, hence the
average $Q$ and $R$ remain essentially the same. However, because in
this case the influence of pairs of particles from different
subsources increases, the effective $\lambda = C_2(0)-1$ now
decreases substantially (as was already observed in \cite{BSWW}). It
should be mentioned at this point that our "Indep" type sources can
probably be used as a possible explanation of the so called
inter-$W$ BEC problem, i.e., the fact that essentially no BEC is
being observed between pions originating from a different $W$ in
fully $W^+W^-$ final states \cite{WW}. It can be understood by
assuming that produced $W$'s should be treated as "Indep" type
sources for which $\lambda$ falls dramatically. Finally, we would
like to stress that our algorithm leads to strong intermittency
showing up after its application. It means that with such algorithm
(which is very efficient and fast for all multiplicities) we can
already attempt to fit experimental data by applying it to some more 
sophisticated fragmentation schemes than that provided by CAS. This
will be done elsewhere. \\ 

\noindent
Acknowledgements:\\
The partial support of Polish Committee for Scientific Research
(grants 2P03B 011 18, 5P03B 091 21 and 621/ E-78/ SPUB/
CERN/P-03/DZ4/99) is acknowledged. The fruitful and stimulating
discussions with B.Andersson, K.Fia\l kowski, T.Cs\"org\H{o} W.Kittel
and S.Todorova-Nova, are gratefully acknowledged.

\begin{figure}
\noindent
\begin{center}
\epsfig{file=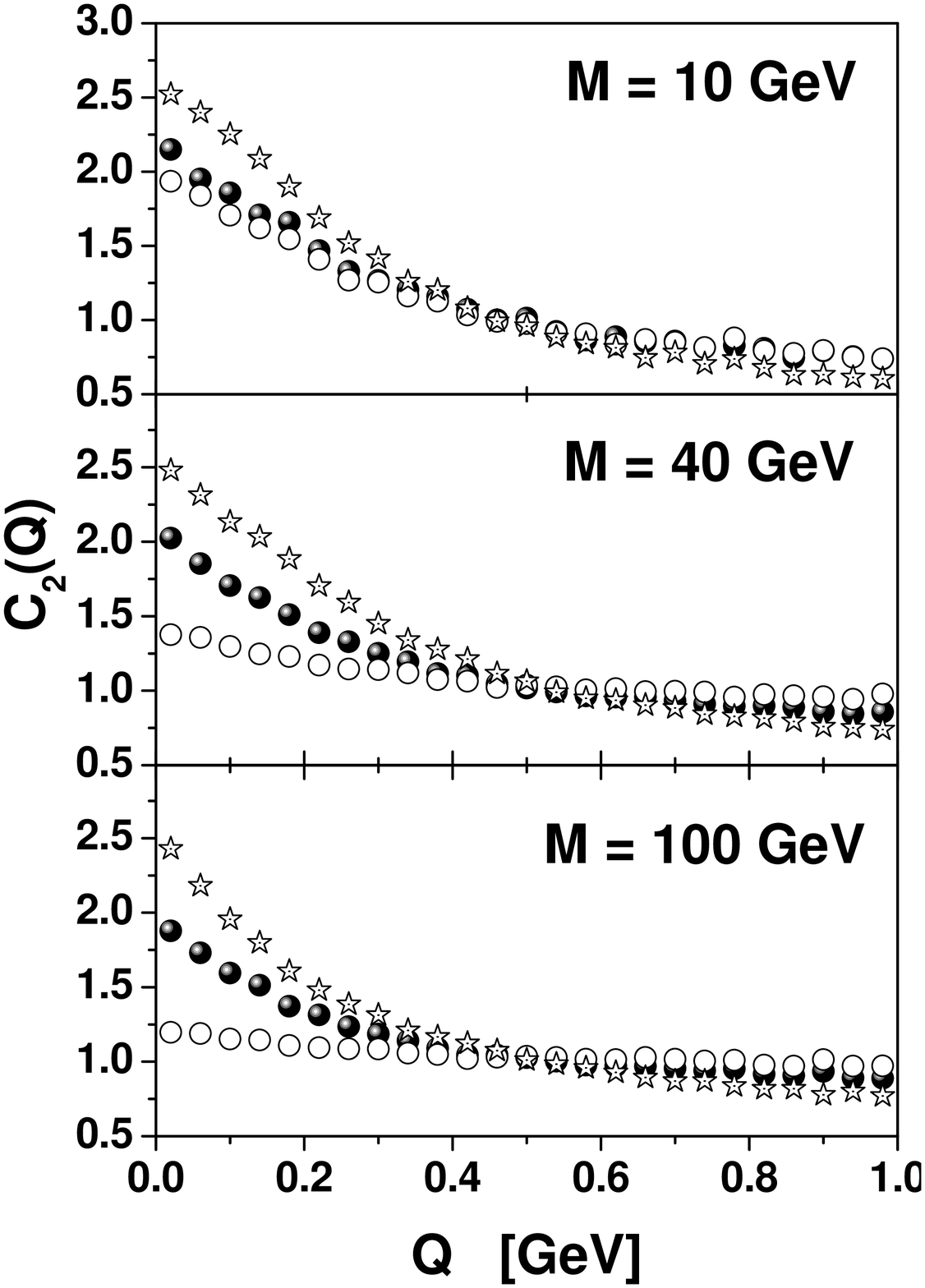, width=120mm}
\end{center}
\caption{Examples of BEC patterns obtained for $M=10$,
$40$ and $100$ GeV for constant weights $P=0.75$ (stars) and $P=0.5$ 
(full symbols) and for the weight given by eq.(\ref{eq:CAS}) 
(open symbols).}
\label{Fig1}
\end{figure}

\begin{figure}[ht]
\noindent
  \begin{minipage}[ht]{70mm}
    \centerline{
        \epsfig{file=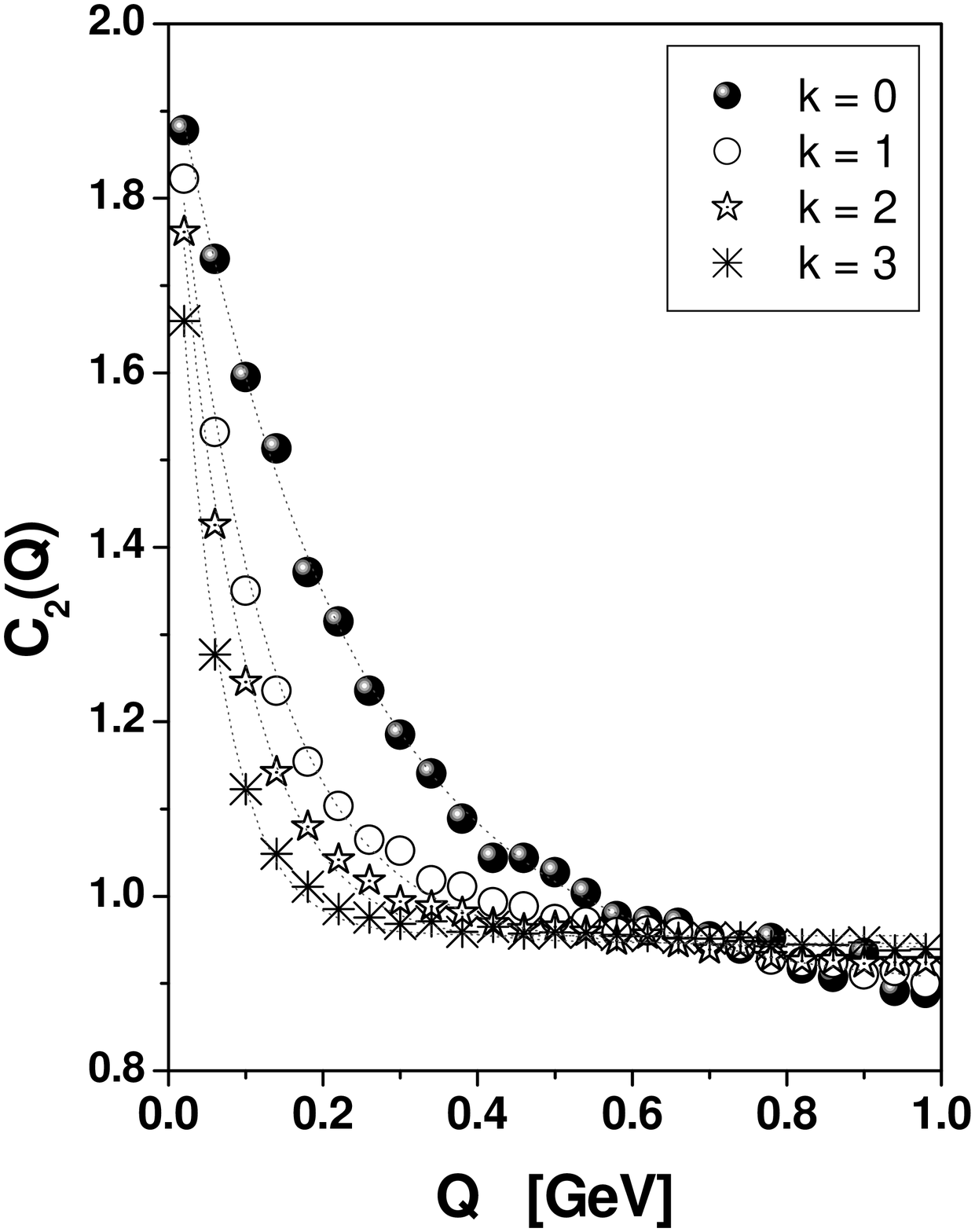, width=70mm}
     }
  \end{minipage}
\hfill
  \begin{minipage}[ht]{70mm}
    \centerline{
       \epsfig{file=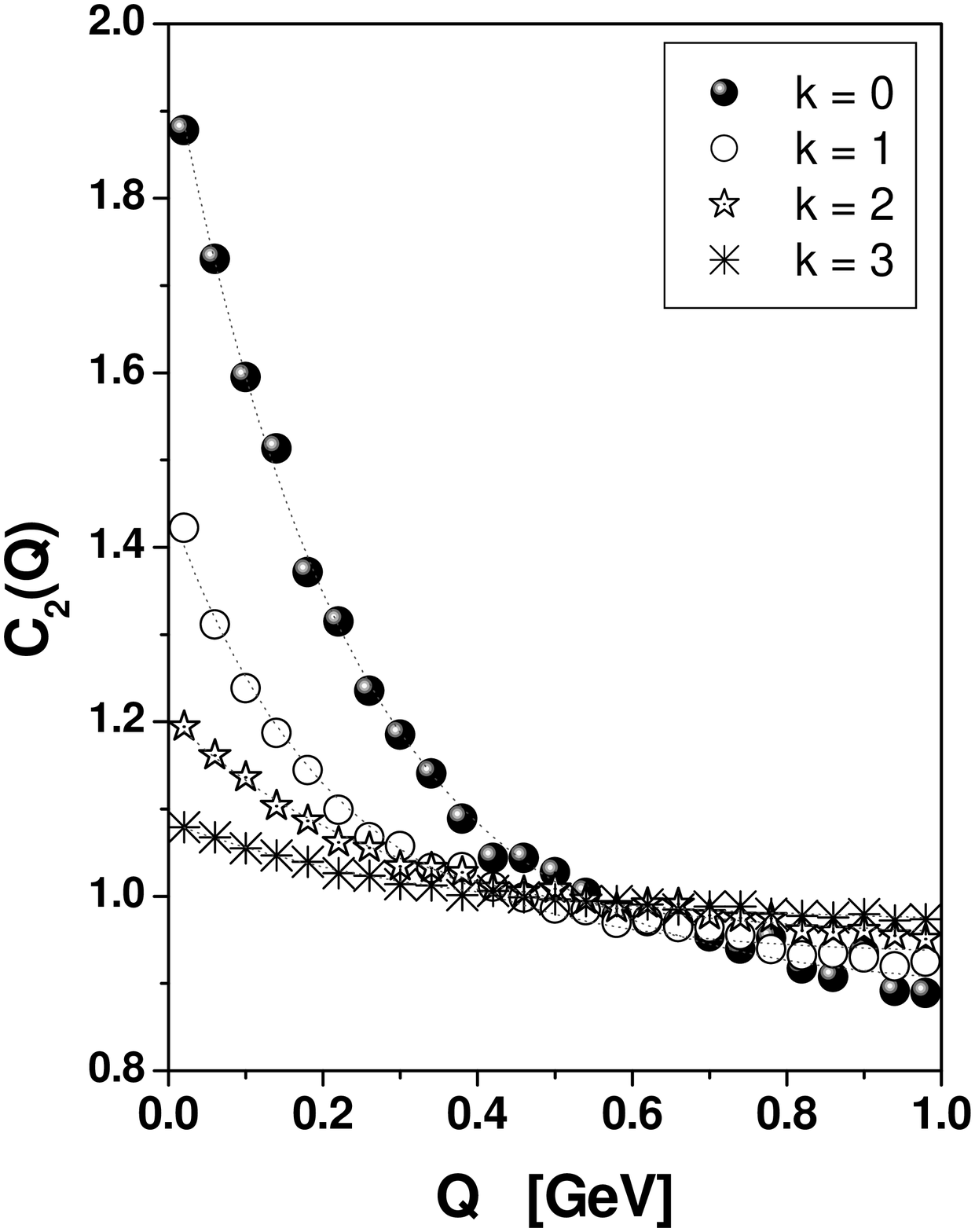, width=70mm}
     }
  \end{minipage}
  \caption{Examples of BEC for different number of
subsources ($n_l=2^k$, $k=1,2,3$ existing in the source $M=100$ GeV.
Left panel is for the "Split" and right panel for the "Indep" types
of sources, as discussed in text.}
  \label{Fig2}
\end{figure}

\begin{figure}[ht]
\noindent
  \begin{minipage}[ht]{70mm}
    \centerline{
        \epsfig{file=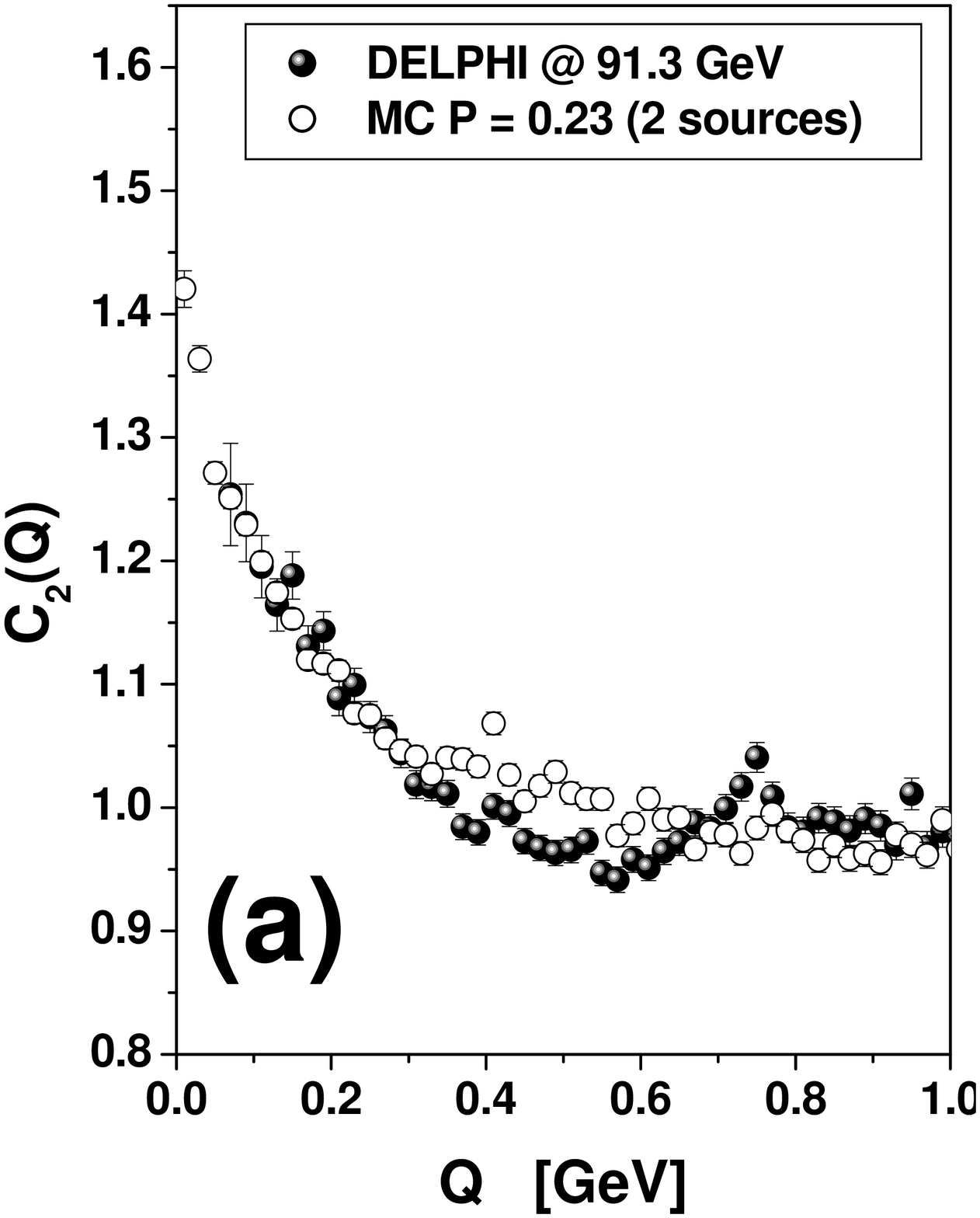, width=70mm}
     }
  \end{minipage}
\hfill
  \begin{minipage}[ht]{70mm}
    \centerline{
       \epsfig{file=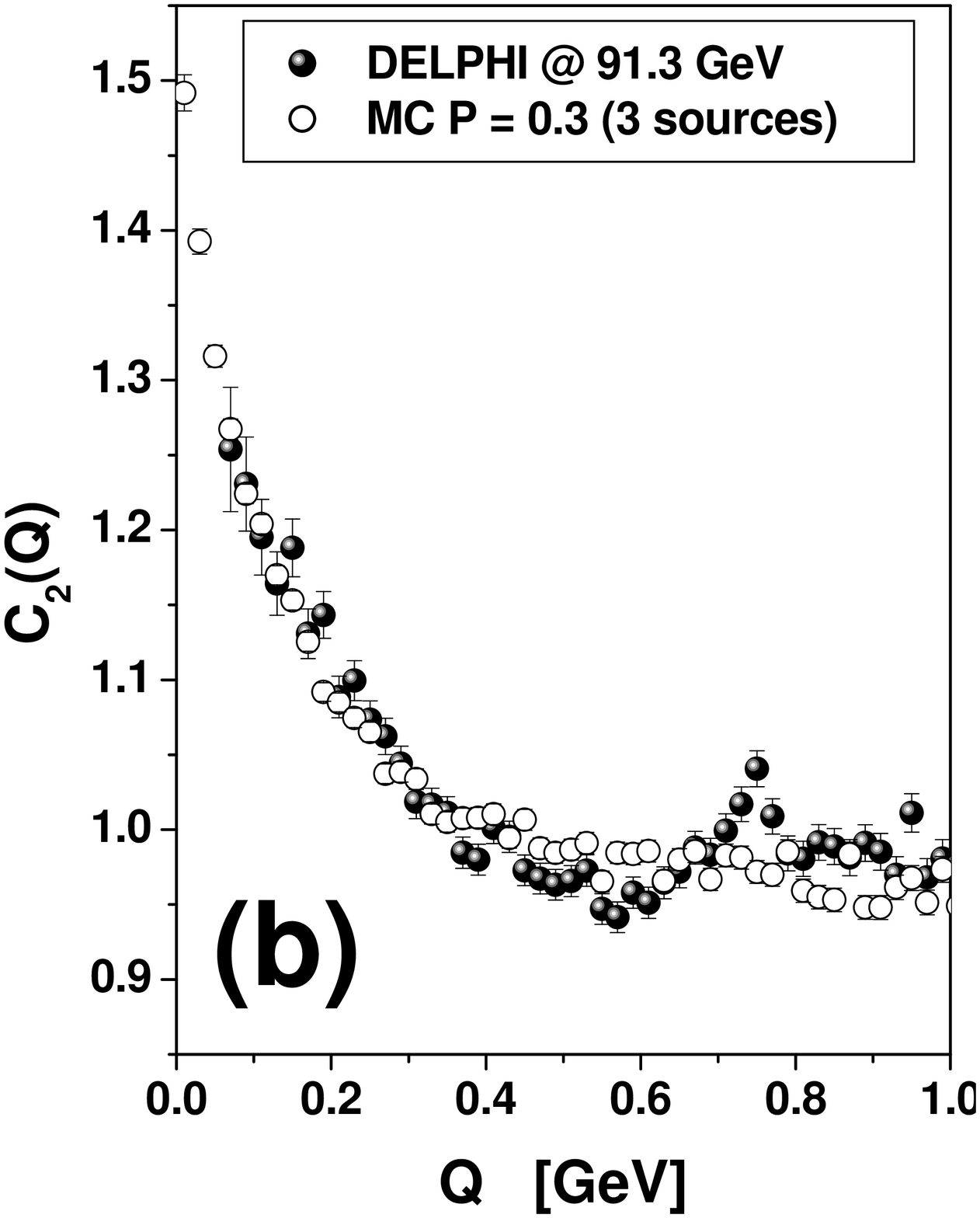, width=70mm}
     }
  \end{minipage}
  \caption{The examples of the "best fits" to the $e^+e^-$
annihilation data on BEC by DELPHI \protect{\cite{DBEC}}
using $2$ $(a)$ or $3$ $(b)$ subsources with $P=0.23$ and 
$P=0.3$, respectively.}
  \label{Fig3}
\end{figure}

\end{document}